\documentclass[preprint,12pt]{elsarticle}

\usepackage{multicol}
\usepackage{graphicx}
\usepackage{booktabs}
\usepackage{amssymb,bm,mathrsfs,bbm,amscd}
\usepackage[tbtags]{amsmath}
\usepackage{lastpage}
\usepackage{verbatim}
\usepackage{comment}
\usepackage{rotfloat}

\journal{Journal of Physics G: nuclear and particle physics}

\begin{document}

\begin{frontmatter}



\title{A new MC-based method to evaluate the fission fraction uncertainty at reactor neutrino experiment \tnoteref{label1}}
 \tnotetext[label1]{Corresponding author}
\author{X.B.Ma\corref{cor1}\fnref{label2}}
 \ead{maxb@ncepu.edu.cn}
\author[label2]{R.M.Qiu}
\author[label2]{Y.X.Chen}
\address[label2]{North China Electric Power University, Beijing,102206,China \fnref{label2}}
\begin{abstract}
Uncertainties of fission fraction is an important uncertainty source for the antineutrino flux prediction in a reactor antineutrino experiment. A new MC-based method of evaluating the covariance coefficients between isotopes was proposed. It was found that the covariance coefficients will varying with reactor burnup and which may change from positive to negative because of fissioning balance effect, for example, the covariance coefficient between $^{235}$U and $^{239}$Pu changes from 0.15 to -0.13. Using the equation between fission fraction and atomic density, the consistent of uncertainty of fission fraction and the covariance matrix were obtained. The antineutrino flux uncertainty is 0.55\% which does not vary with reactor burnup, and the new value is about 8.3\% smaller.

\end{abstract}


\end{frontmatter}

\section{Introduction}
Reactor antineutrino is used to study neutrino oscillation, search for signatures of nonstandard neutrino interaction, and monitor reactor operation for safeguard application. Antineutrino flux is an important source of uncertainties for a reactor neutrino experiment.The antineutrino flux and spectrum of $\bar{\nu}_{e}$ from a reactor on a given day t can be predicted by equation

\begin{equation}
\label{total_flux}
S(t) = \int_{E_{\nu}}\frac{W_{th}(t)}{\sum_{i}f_{i}(t)E_{i}}\sum\limits_{i}f_{i}(t)S_{i}(E_{\nu})dE_{\nu}
\end{equation}

where $S(t)$ is the total flux,$L$ is the baseline, $W_{th}$ is the thermal power of the reactor, $f_{i}$, $E_{i}$, $S_{i}(E_{\nu})$ are the fission fraction, energy release and antineutrino spectrum of isotopes i respectively. The uncertainty of antineutrino flux by the fission fraction $\frac{\delta S}{S}$ can be calculated.
\begin{equation}
\label{uncer}
\frac{\delta S}{S}= \frac{1}{S}\sqrt{\sum\limits_{i,j}\frac{\partial S}{\partial \alpha_{i}}\frac{\partial S}{\partial \alpha_{j}}\cdot\delta\alpha_{i}\delta\alpha_{j}\rho_{i,j}}
\end{equation}
Where $\delta\alpha_{i}$ is the uncertainty of fission fraction of the isotopes $i$ ($i$=$^{235}$U,$^{238}$U, $^{239}$Pu and $^{241}$Pu), $\rho_{i,j}$ are the correlation coefficients between isotopes.

During the power cycle of a nuclear reactor, the composition of the fuel changes as Pu isotopes are bred and U isotopes are depleted. At the end of the power cycle, some fraction of the fuel is replaced, for example, Daya Bay reactor, one third of fuel will be replaced. The fresh fuel always stay in the reactor for three cycles in order to generate more power. Knowledge of the detailed time-depended of the content of the fuel is of interest to the reactor operator and designer for the safety consideration. To obtain this knowledge, reactor simulation code were developed such as DRAGON\cite{DRAGON}, Reactor Monte Carlo code (RMC)\cite{RMC}, CASMO,SCALE\cite{SCALE} and so on. The verification and validation of the code simulation was done by comparison of isotropic concentration with the experiment results. The fission fraction coefficients between isotopes\cite{Djurcic} was approximately studied using 159 comparisons of fuel element samples taken from ten PWRs and BWRs, modeled by a variety of core simulation codes, because these isotopic concentration comparisons only give indirect information
on the uncertainty in the number of fissions fraction. The correlation coefficients was also studied by using fission fraction of each isotopes directly\cite{fissionmaxb}. However, for the next generation reactor antineutrino experiment (JUNO) which is aiming to perform high precision neutrino oscillation measurements, it is necessary to known more information about the uncertainty of fission fraction, such as, what is relation between the uncertainty of flux prediction caused by fission fraction and reacor burnup. Previous study can only give the burnup average fission fraction uncertainty, but not the information with burnup. In this study, this question was addressed by using a new MC-based method.

In section \ref{method}, The MC-based method of evaluating the correlation coefficient between different isotopes was introduced. The parameters, which are needed in the method, such as atomic density and one group microscopic cross sections as a function of burnup, are calculated by RMC. The model of this calculation was discussed in section \ref{reactor}. The covariance matrix results were discussed in section \ref{results}, and its effect to the uncertainty of the antineutrino flux was discussed in section \ref{burnup},and the last section is the conclusion.

\section{correlation coefficient evaluation method}
\label{method}

Generally, the reaction rate can be defined as
\begin{equation}\label{rate0}
  R^{i}_{f} = \Sigma^{i}_{f} \times \bar{\phi} = N_{i}\bar{\sigma}^{i}_{f}\bar{\phi}
\end{equation}
where, $R^{i}_{f}$ is the fission rate of isotopes $i$, $\Sigma^{i}_{f}$ is the average macroscopic fission cross section, $N_{i}$ is the atomic density of isotopes $i$, $\bar{\sigma}^{i}_{f}$ is the average microscopic fission cross section, and $\bar{\phi}$ is the average neutron flux. The fission fraction can be defined as
\begin{equation}\label{rate1}
  f_{i} = R^{i}_{f}/\sum_{j}(R^{j}_{f})=N_{i}\bar{\sigma}^{i}_{f}/\sum_{j=1}^{4}(N_{j}\bar{\sigma}^{j}_{f})
\end{equation}

\begin{equation}
\label{ccf}
  \rho_{i,j}=\frac{1}{N-1}\sum_{k=1}^{N}\frac{(f_{i}-\bar{f_{i}})}{\sigma_{f_{i}}}\frac{(f_{j}-\bar{f_{j}})}{\sigma_{f_{j}}}
\end{equation}
where $N$ is the total sample number, $\sigma_{f_{i}}$ and $\sigma_{f_{j}}$ are the standard deviation of $f_{i}$ and $f_{j}$ respectively. To suing this method, the difficulty is how to obtain the one group average cross section and atomic density of each isotopes as a function of burnup accurately. This question was discussed in next section.

\section{Reactor Simulation and sampling}
\label{reactor}
\begin{figure}
\begin{center}
\includegraphics[width=9cm]{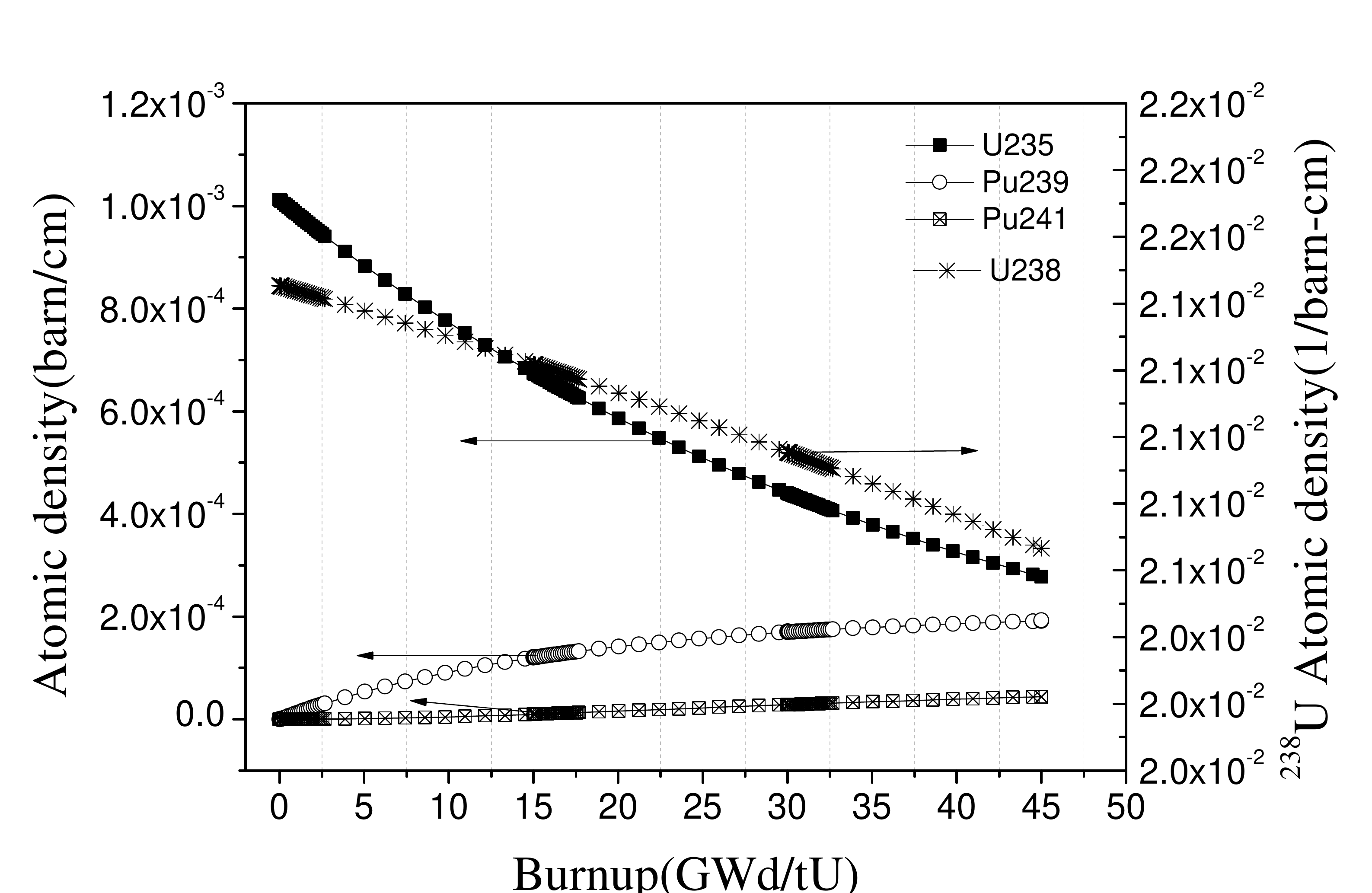}
\caption{The atomic density as a function of burnup}
\label{atomicdensity}
\end{center}
\end{figure}

\begin{figure}
\begin{center}
\includegraphics[width=9cm]{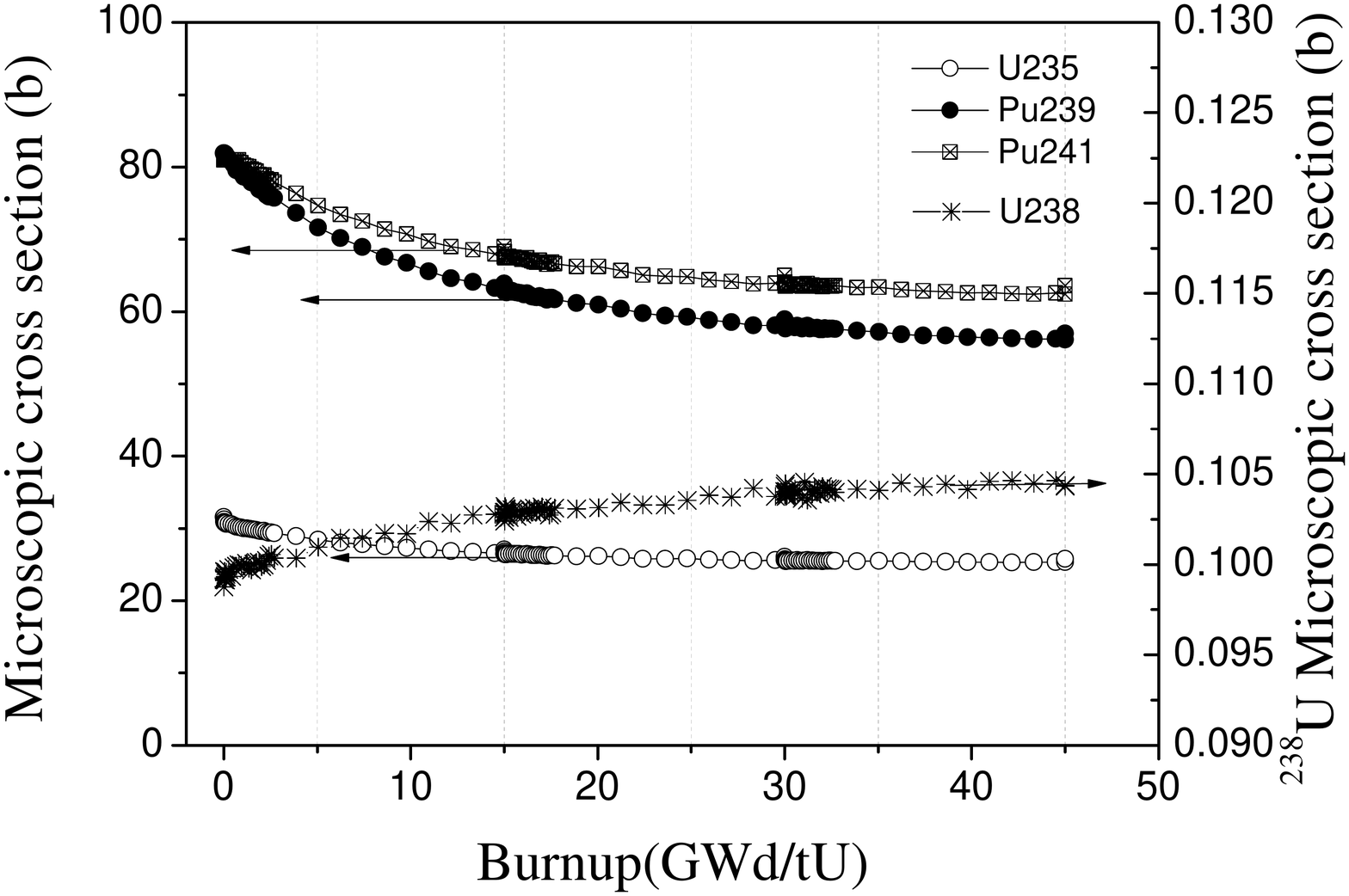}
\caption{The one group average cross section as a function of burnup}
\label{xs-burnup}
\end{center}
\end{figure}

   To evaluate the fission fraction from equation (\ref{rate1}), the atomic density and the average cross sections of each isotopes must be obtained. Reactor simulation code is necessary to use because of strong energy dependence of cross section, angular dependence severe locally, temperature dependence of cross sections. The Daya Bay reactor operates with 157 fuel assemblies producing a total thermal power of 2895MW. The assembly is a 17$\times$17 design, a total of 289 rods. There are 264 fuel rods, 24 control rods and one guide tub. The enrichment of new fuel is 4.45\% for the 18-months reloaded design. An ordinary PWR pin cell with fuel enrichment of 4.45\% was made and RMC code was used to simulate the atomic density and one group average cross section of each isotopes. Fig.\ref{atomicdensity} shows the atomic density as a function burnup. The atomic densities of $^{235}$U and $^{238}$U are decreasing with the burnup increasing mainly because of fission and capture reactions. On the other hands, the atomic densities of $^{239}$Pu and $^{241}$Pu are increasing with the burnup increasing mainly because of transmutation from $^{238}$U. Fig.\ref{xs-burnup} shows the one group average cross sections of each isotopes as a function of burnup. The one group average cross sections of $^{238}$U is increasing with the burnup increasing because that the fission cross sections $^{238}$U is increasing with the energy increasing, and the neutron spectrum becoming more harder at the end of three cycles, which causes the average of the neutron increasing. The one group average cross sections of $^{235}$U, $^{239}$Pu and $^{241}$Pu are decreasing with the burnup increasing because the fission cross sections of above two isotopes decreasing with neutron energy increasing.

\begin{figure}
\begin{center}
\includegraphics[width=6cm]{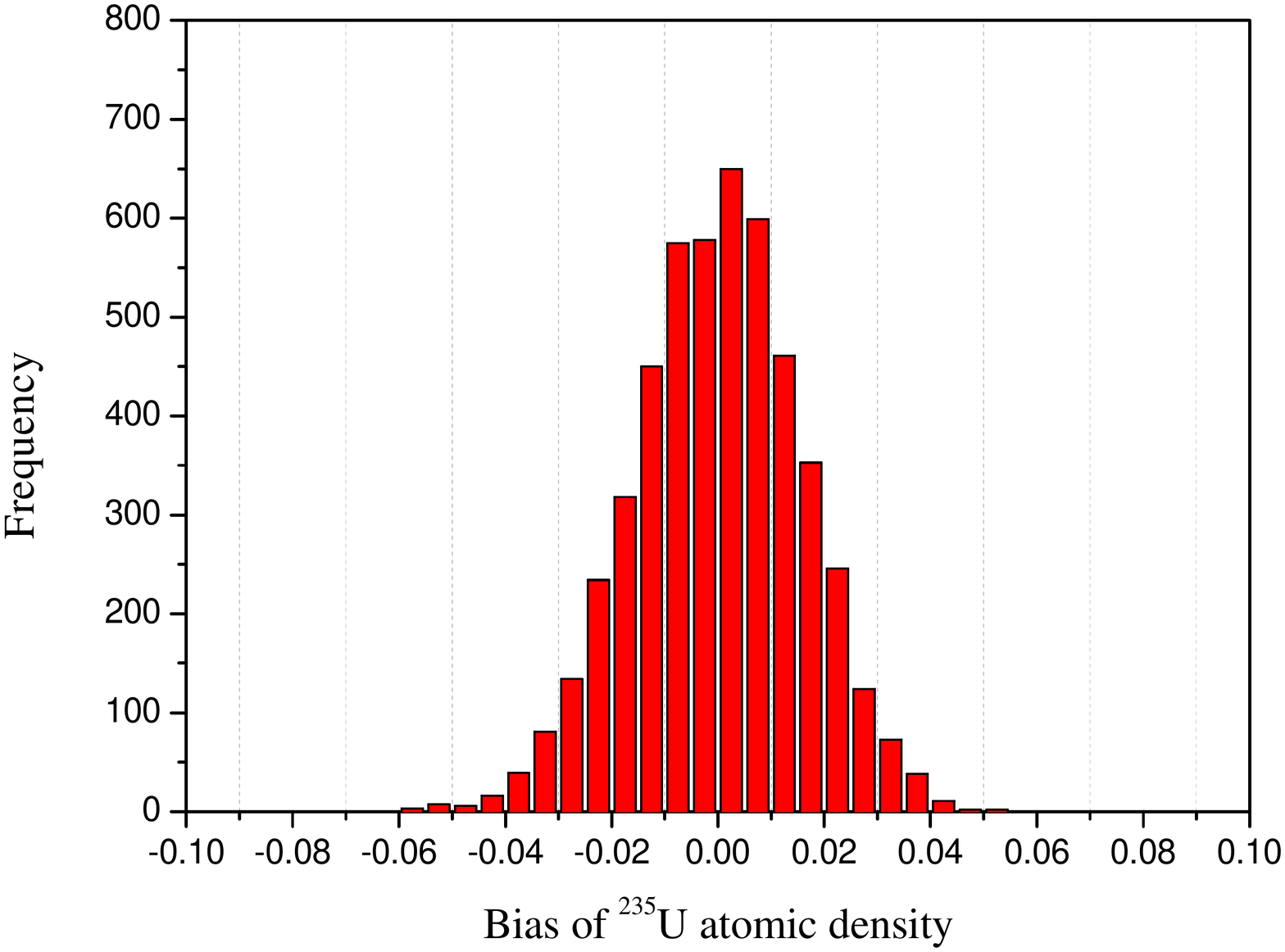}
\includegraphics[width=6cm]{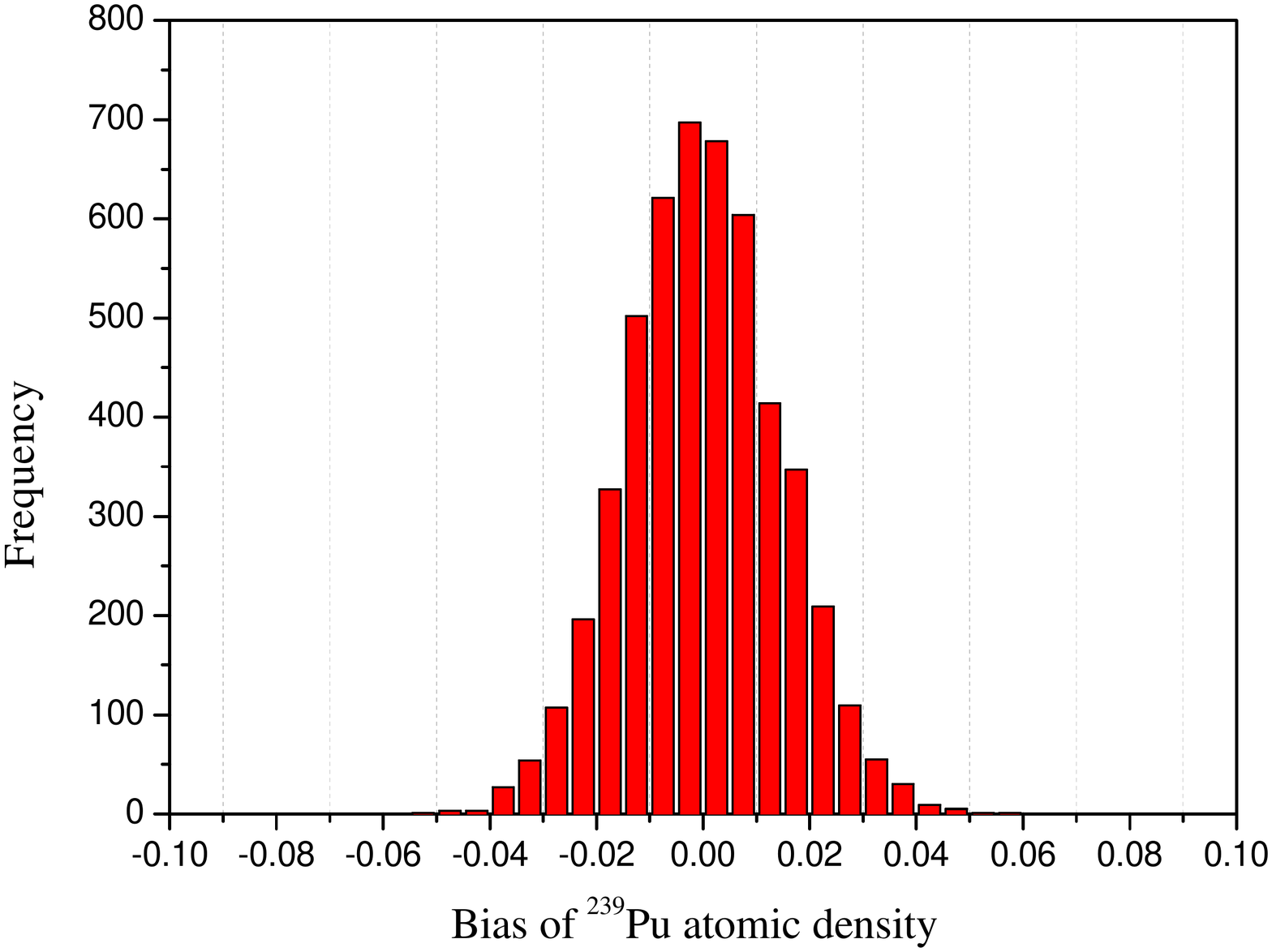}
\caption{The atomic density of $^{235}$U and $^{239}$Pu distribution}
\label{AtomicDensity1}
\end{center}
\end{figure}

   To obtain the sample of the fission fraction, the distributions of variables $N_{i}$ and $\sigma^{i}$ should be known. It is reasonable to assume that the distributions of $N_{i}$ and $\sigma^{i}$ are standard distribution.
   The atomic density difference of each isotopes ($^{235}$U,$^{238}$U,$^{239}$Pu and $^{241}$Pu) between calculation and measurement were about 5\%\cite{fissionmaxb}, therefor, the value 5\% was used as the standard deviation of atomic density of each isotopes.
   Macroscopic fission cross section uncertainty analysis were studied using TSUNAMI-2D sequence in SCALE6.1. It was found that the fission cross section uncertainty of PWR is less than 0.36\%\cite{u1}. However, the statical uncertainty of the one group average cross section of each isotopes is less than 0.1\% when doing the simulation using RMC. Considering the conservative of the results, the value 0.36\% was adopted as the standard deviation of the one group average cross section of each isotopes. Fig.\ref{AtomicDensity1} shows the atomic density of $^{235}$U and $^{239}$Pu distribution as standard normal distribution, and which demonstrate that the sample code work correctly.

\section{Correlation coefficient results}
\label{results}
\begin{figure}
\begin{center}
\includegraphics[width=6cm]{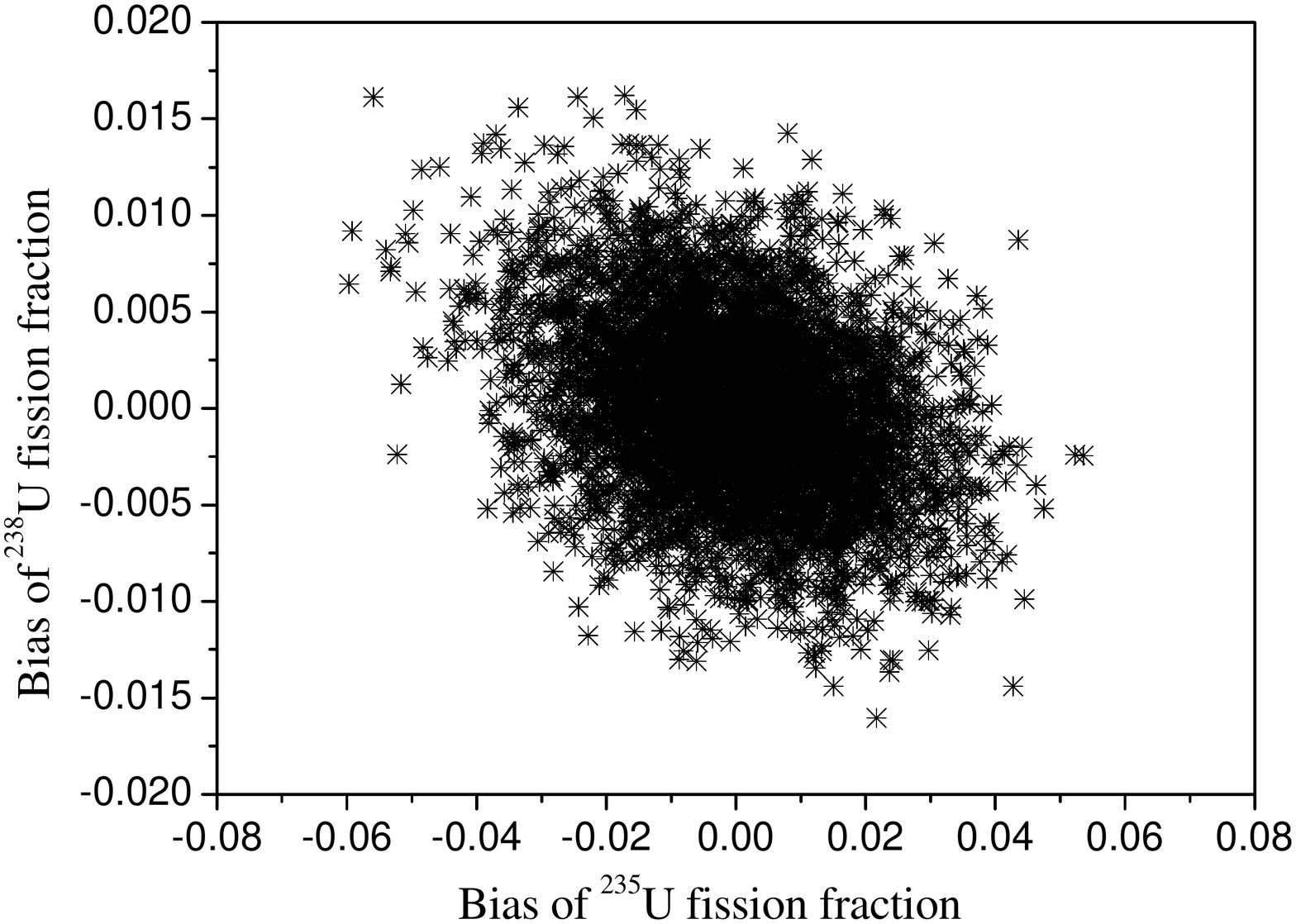}
\includegraphics[width=6cm]{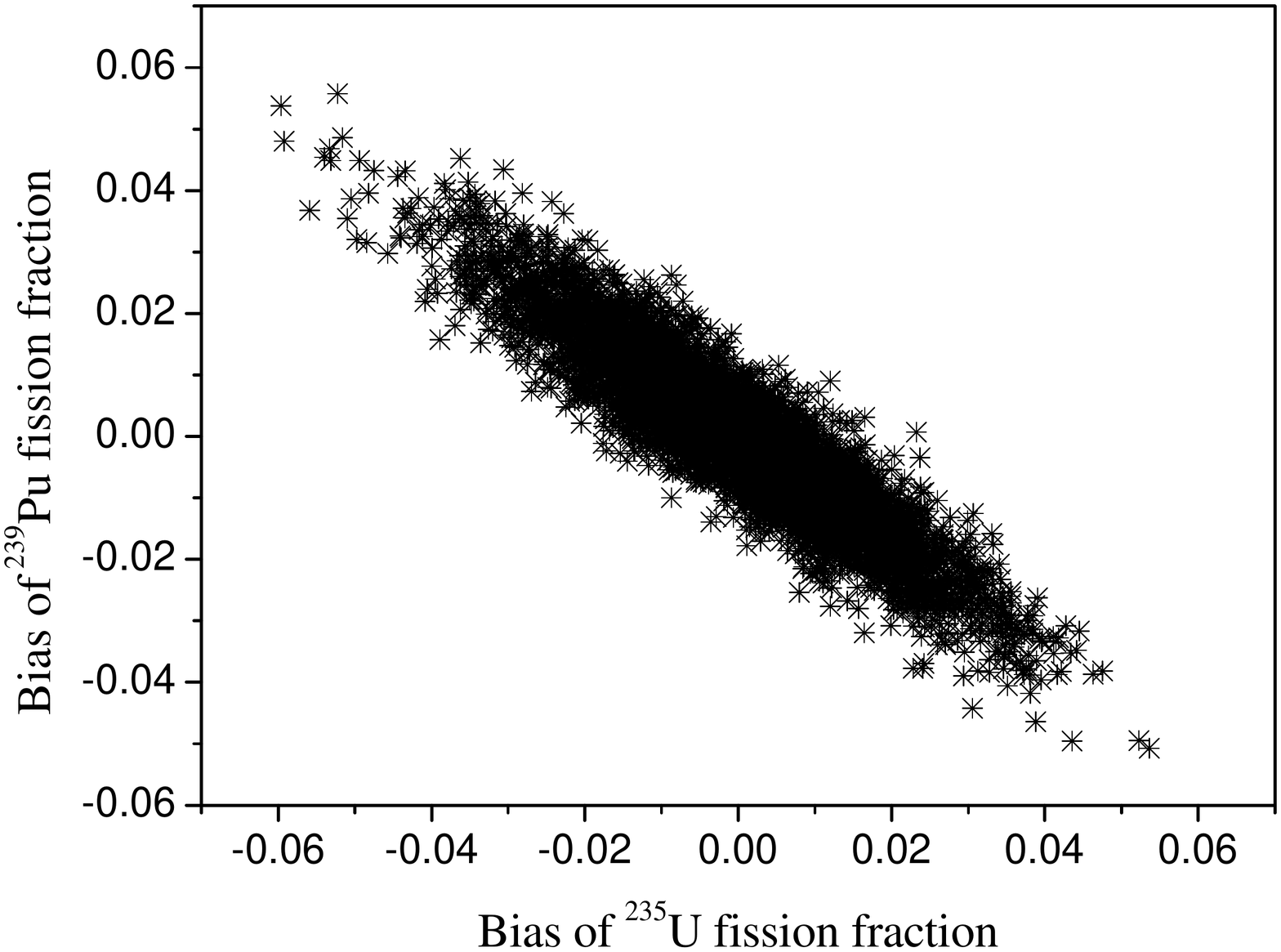}
\includegraphics[width=6cm]{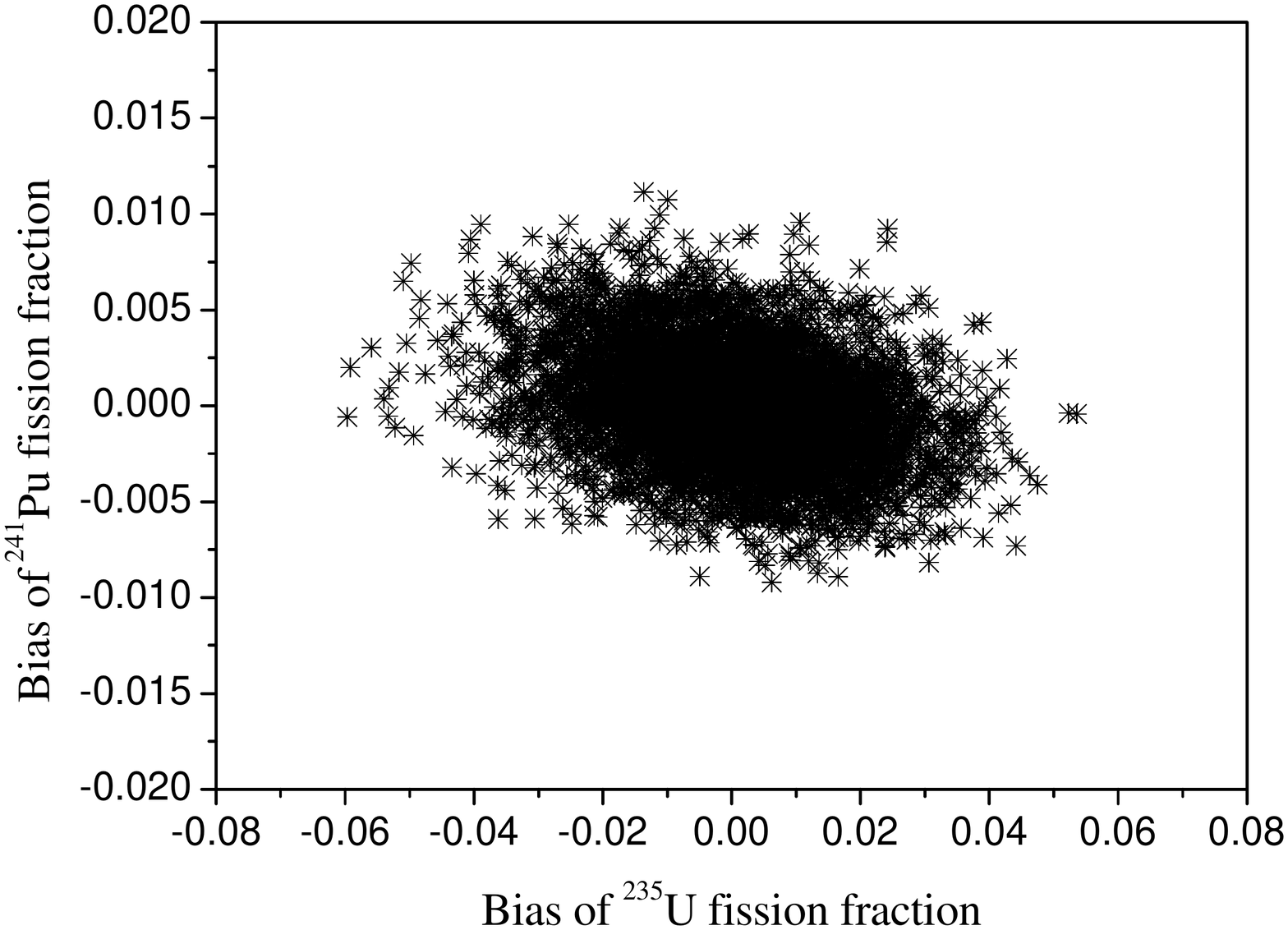}
\includegraphics[width=6cm]{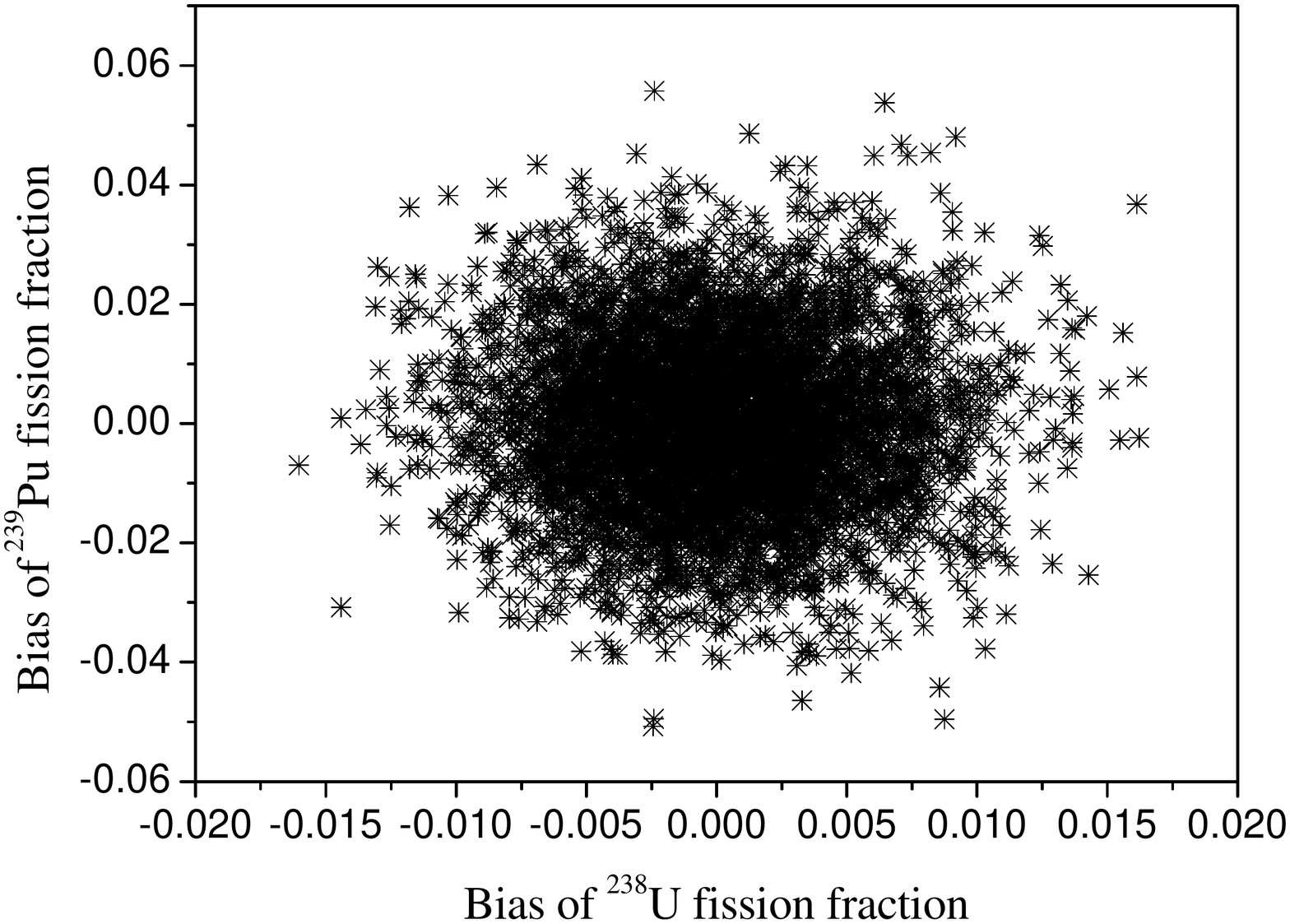}
\includegraphics[width=6cm]{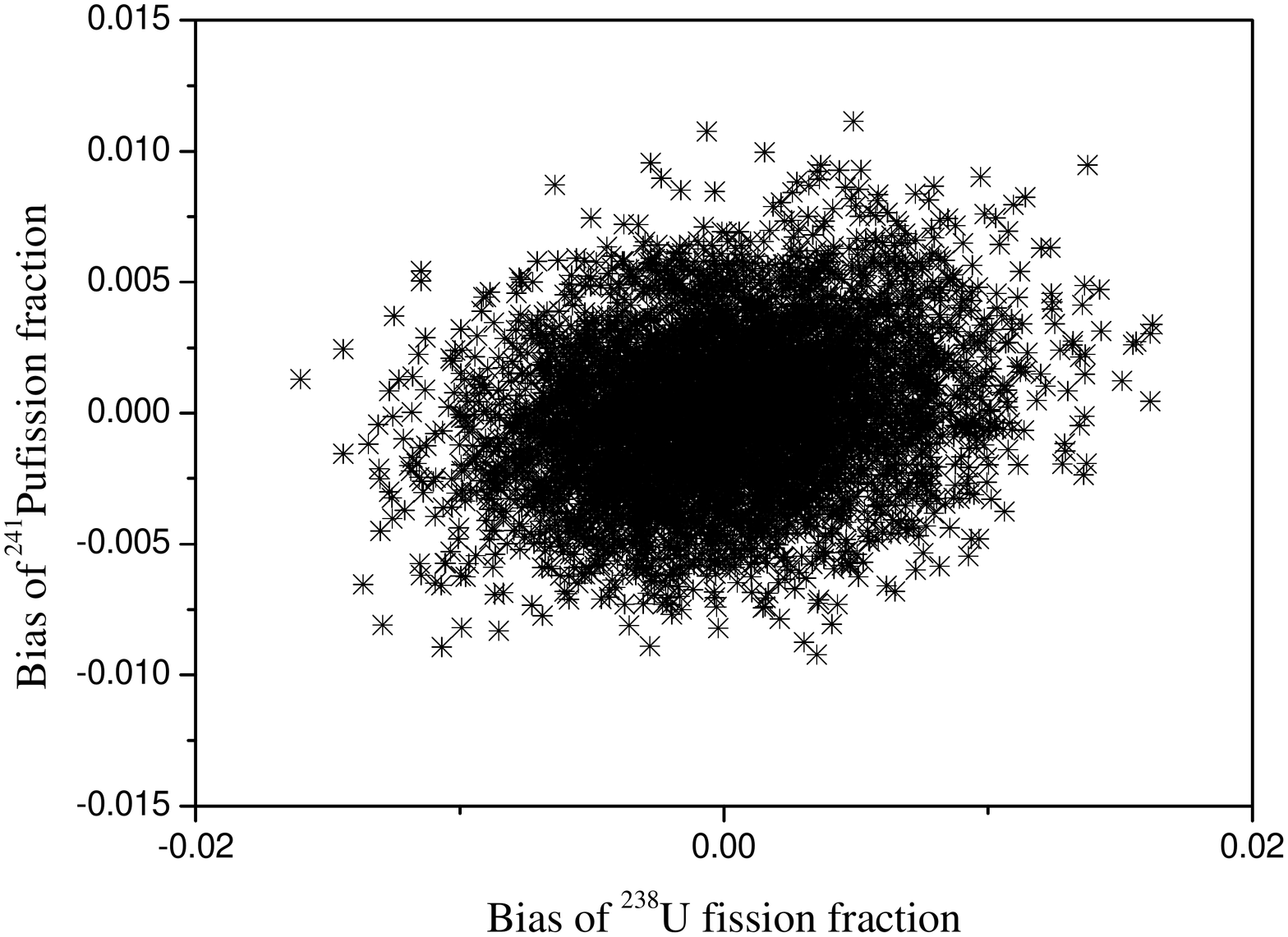}
\includegraphics[width=6cm]{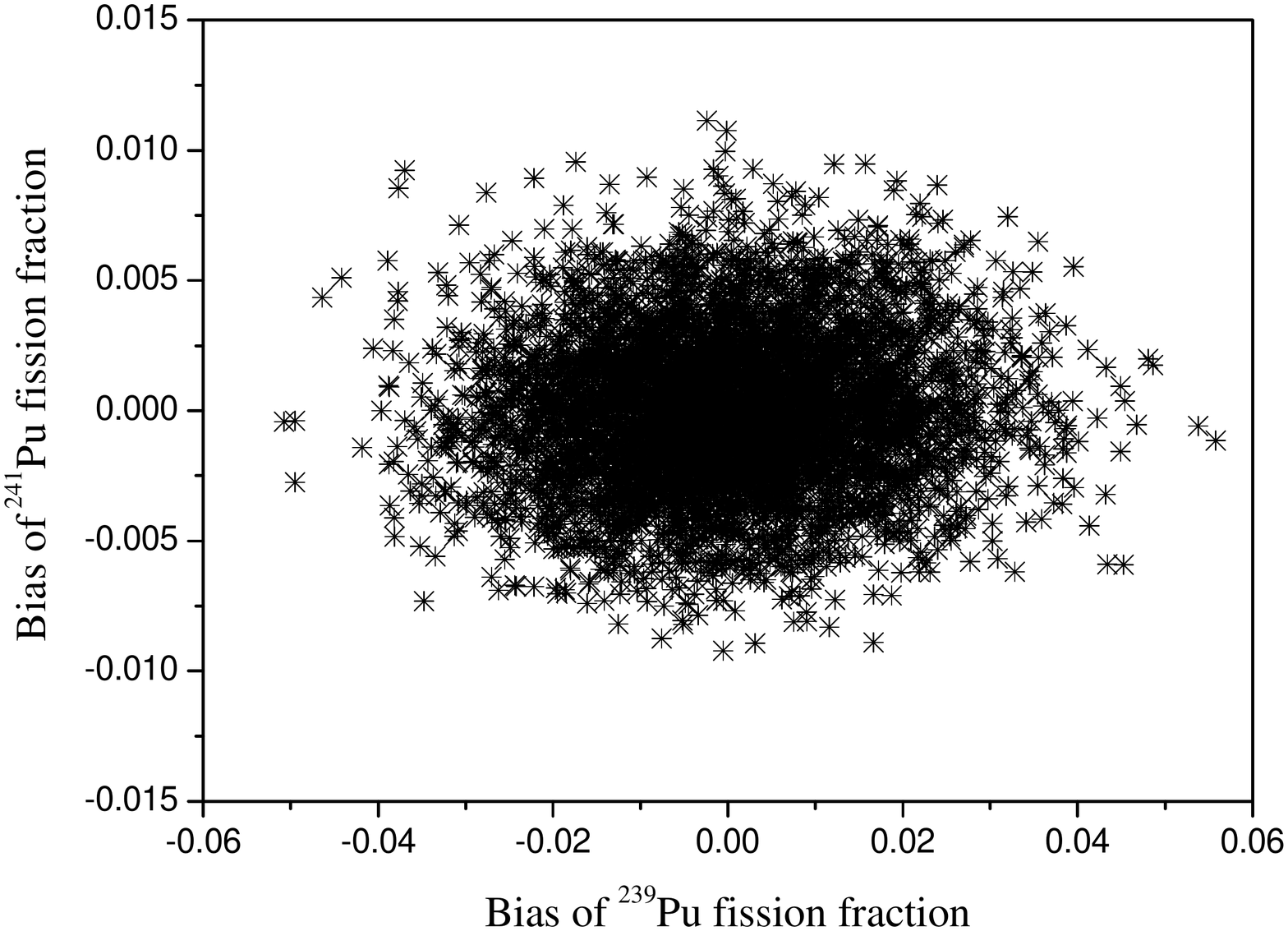}
\caption{The activities in one batch of long-life fission products as a function of time after shutdown}
\label{correlations}
\end{center}
\end{figure}

  Using equation (\ref{ccf}), correlations between the $f_{i}$ were studied as well. Fig.(\ref{correlations}) plots the $f_{i}$ for pairs of fuel isotopes for 50 thousands samples comparisons at the begin of cycle. We find a weak anti-correlation between $^{235}$U and $^{238}$U and $^{241}$Pu, but a strong anti-correlation between $^{235}$U and $^{239}$Pu. This is expected since $^{235}$U and $^{239}$Pu dominate the thermal power production of the reactor, and so to maintain the energy balance, the over fissioning of this isotope must be accompanied by an under-fissioning of the other isotopes (and visce-versa). The fact that the largest anti-correlation is with $^{239}$Pu, strengthens this arguments.

  To evaluate the correlation coefficients as a function of burnup, we assume that the standard deviation of the atomic density of each isotopes and one group average fission cross section are not varying with burnup, just the atomic density of each isotopes and one group average fission cross section are varying with burnup. The correlation coefficients as a function of burnup were evaluated as shown in Table 1. The correlation coefficients are varying with burnup linearly. On the other hands, some correlation coefficients may change its sign, for example, the correlation coefficients of ($^{238}$U, $^{239}$Pu) and ($^{239}$Pu, $^{241}$Pu) changes from positive to negative. There are two reasons which would cause that results. One is decay chain, for example, the correlation coefficients of ($^{238}$U, $^{239}$Pu) would be a positive because that $^{238}$U will be transmutation to $^{239}$Pu after absorption one neutron and then decaying two times. This effect could be called positive effect. Otherwise, the over fissioning of $^{239}$Pu must be accompanied by an under-fissioning of the other isotopes which include $^{238}$U, and which would be an negative effect. The results the correlation coefficient is positive or negative effect can be explained by which effect will dominate the process. For instance, at the begin of cycle, the positive effect is more important than the negative effect, the correlation coefficients of ($^{238}$U, $^{239}$Pu) is positive. But, at the end of cycle, the situation is the opposite.

\begin{table}
\label{activity}
\caption{Covariance matrix of different isotopes vs reactor burnup}
\begin{center}
\footnotesize
\begin{tabular*}{155mm}{ccccccc}
\toprule
Burnup(GW.d/tU) & $^{235}$U-$^{238}$U & $^{235}$U-$^{239}$Pu& $^{235}$U-$^{241}$Pu& $^{238}$U-$^{239}$Pu&
$^{238}$U-$^{241}$Pu &	$^{239}$Pu-$^{241}$Pu \\
  \hline
0    & -0.48& -0.93& -0.41& 0.15& 0.30& 0.22\\
1.07 & -0.45& -0.93& -0.40& 0.12& 0.28& 0.19\\
2.02 & -0.43& -0.93& -0.38& 0.10& 0.28&	0.17 \\
5.05 & -0.36& -0.92& -0.33&	0.026& 0.25& 0.10 \\
7.42 & -0.32& -0.92& -0.31& -0.019& 0.23 & 0.057 \\
10.96& -0.26& -0.92& -0.27& -0.082& 0.21 & -0.010 \\
13.33& -0.23& -0.91& -0.25& -0.12 &0.19 & -0.056\\
15.00& -0.22& -0.91& -0.24& -0.13 & 0.19& -0.065\\
\bottomrule
\end{tabular*}
\end{center}
\end{table}

\section{Correspondence between mass inventory and fission rates}
\label{corre}
   The mass inventory was calculated and was comparison with the measurement data\cite{fissionmaxb} \cite{APOLLO}, and $^{235}$U, $^{239}$Pu and $^{241}$Pu was predicted with an experimental uncertainty below 5\%($1\sigma$). From equation \ref{rate1}, the mass inventory uncertainty did not equal to the fission fraction uncertainty. In the study, the fission fraction uncertainty was calculated by assumption that the uncertainty of the atomic density of each isotopes was 5\% and
   the correlation between isotopes was also considered. The uncertainty of fission fraction of different isotopes was shown in table \ref{mass}.

\begin{table}
\caption{Uncertainty of fission fraction of different isotopes(\%)}
\begin{center}
\footnotesize
\begin{tabular*}{60mm}{ccccc}
\toprule 	& $^{235}$U & $^{238}$U & $^{239}$Pu &$^{241}$Pu \\
  \hline
BOC	&2.03 & 5.84 & 5.19& 6.08 \\
MOC	&2.80 & 5.63 & 4.41& 5.76 \\
EOC &3.30 & 5.50 & 3.98& 5.60 \\

\bottomrule
\end{tabular*}
\label{mass}
\end{center}

\end{table}

\section{Uncertainty of antineutrino flux with burnup}
\label{burnup}
  To quantify the effect of the new values for correlation coefficients on antineutrino flux expectation in a reactor neutrino experiment. The reactor data from the Daya Bay experiment were used to calculate the expected average daily antineutrino flux at one detector. Due to the covariance matrix varying with reactor burnup, the uncertainty of fission fraction will also varies with burnup, as shown in table \ref{resultsburnup}. Comparison the results with the previous average burnup (Djurcic and Xubo Ma's results), the uncertainty is a little larger. It is a increasing with the burnup increasing because of the correlation coefficients may change its sign with the burnup increasing. In above case, the fission fraction uncertainty of all isotopes were assumed as the uncertainty of mass inventory 5\%, which was found in Ref.\cite{fissionmaxb} \cite{APOLLO}. However, in section \ref{corre}, it was found that the uncertainty of mass inventory was not same as fission fraction. If the uncertainty of each isotopes in table \ref{mass} was used to evaluate the uncertainty of antineutrino flux as a function of burnup, the uncertainty of antineutrino flux was 5.5\% and which did not vary with reactor burnp because the consistent between the new uncertainty of fission fraction and the new covariance matrix.

\begin{table}

\caption{Uncertainty of neutrino flux result by fission raction using different covariance matrix}
\begin{center}
\footnotesize
\begin{tabular*}{95mm}{cc}

\toprule Type of covariance matrix  & Uncertainty of fission fraction \\
  \hline
  Begin of Cycle  & 0.65\% \\
  Middle of Cycle & 0.68\% \\
  End of Cycle    & 0.70\% \\
  Djurcic et al\cite{Djurcic}.  & 0.58\% \\
  Xubo Ma et al\cite{fissionmaxb}   & 0.60\% \\
\bottomrule
\end{tabular*}
\label{resultsburnup}
\end{center}

\end{table}

\section{Conclusion}
   A new MC-based method of evaluating the covariance matrix between isotopes was proposed. The method has the capability of considering the uncertainty of atomic density and microscopic cross section at the same time and evaluating the covariance coefficient varying with burnup. We find that the covariance coefficient may change from positive to negative because of fissioning balance effects. This fact would cause the uncertainty of fission fraction having a little larger. Using the equation between fission fraction and atomic density, the consistent of uncertainty of fission fraction and the covariance matrix were studied. The antineutrino flux uncertainty is 0.55\% which does not vary with reactor burnup, and the new value is about 8.3\% smaller.

\section*{Acknowledgements}
The work was supported by National Natural Science Foundation of China (No.11390383) and the Fundamental Research Funds for the Central Universities (2015ZZD12), and Cao Jun for his extraordinary support.





\end{document}